\documentclass{article}
\usepackage{fleqn,ams}
\usepackage{amssymb}
\usepackage{epsfig}
\usepackage{verbatim}
\def\stackunder#1#2{\mathrel{\mathop{#2}\limits_{#1}}}
\newcommand{\REFigure}[2]{%
\begin{center}\epsfig{file=#1,width=8cm,height=8cm,angle=-90}\\[12pt]
\refstepcounter{figure}Figure \thefigure: {\sl #2}
\end{center}}
\begin{document}

\begin{center}
{\bf\Large  The Kinetics Of Nonequilibrium Universe. I. The
Condition Of Local Thermodynamical Equilibrium.}\\
Yu.G.Ignatyev\\
Kazan State Pedagogical University,\\ Mezhlauk str., 1, Kazan
420021, Russia
\end{center}

\begin{abstract}
In terms of fundamental principles of quantum theory of
interacting particles and relativistic kinetic theory there was
carried out an analysis of the main principle of standard
cosmological scenario - the initial existence of local
thermodynamical equilibrium. It has been shown, that condition of
the existence of local thermodynamical equilibrium in Universe is
determined essentially by means of function of total cross-section
of particles' interaction from the kinematic invariant and in case
of scaling's recovery in range of superhigh energies it is
initially broken.
\end{abstract}


\section{Standard View On The Establisment Of LTE In Universe}
The most prevalent, established from the very first papers on
cosmology, point of view on the problem of local thermodynamic
equilibrium (LTE) in early Universe, is posed in well-known
J.B.Zeldovich's and I.D.Novikov's monograph \cite{Zeld}, (1975):

``...As it was mentioned above, all particles lay in
thermodynamical equilibrium at high temperature. Actually, for the
existence of the thermodynamical equilibrium it is required that
processes, establishing the equilibrium have to run faster, than
plasma's establishment. More precisely, it is necessary, that
during the period of process, which establishes the equilibrium
($\tau$), there were far less of character time of plasma
parameters' variation ($\rho, T$ etc).

In isotropic solution $\rho={\displaystyle \frac{\alpha}{Gt^2}}$,
where $\alpha$ is of one order. Therefore time, essential for the
variation of density from some value $\rho$ up to
$\left({\displaystyle \frac{1}{e}}\right)\rho\approx 0,4\rho$, ïîðÿäêà %
$\Delta t={\displaystyle \frac{1}{\sqrt{G\rho}}}$.

Thus, $\Delta t$ is of order $t$. From the other hand,
equilibrium's establishment time is
$$
\tau=\frac{1}{\sigma nv},\eqno{(6.2.6)}
$$
where $\sigma$ - reaction's cross-section, $n$ - concentration of
particles, $v$ - velocity of their movement. At high temperatures
$v\approx c$. Value $n$ is determined by formula (6.2.5):
$n=n_1t^{-3/2}$ ($n_1=\mbox{const}$). Therefore
$$
\tau=\frac{t^{3/2}}{\sigma n_1 c}. \eqno{(6.2.7)}
$$
For the thermodynamical equilibrium is necessary:
$$
\tau=\frac{t^{3/2}}{\sigma n_1 c}<\Delta t \approx t
\eqno{(6.2.8)}
$$
or
$$
\sigma>t^{1/2}n_1c. \eqno{(6.2.9)}
$$
Therefore thermodynamical equilibrium exists at $t\to 0$, if only
$\sigma$ does not decrease sufficiently fast together with the
extension of particles' energy. One may hope, that condition
(6.2.9) is fulfilled in fact. Thus, for example, it is beyond any
doubt, that at high temperatures the number of pairs $e^+,e^-$
does not differ from the equilibrium one. In fact, let us consider
as an example the moment, when $T=1$ MeV, $t=1$ sec,
$n_{e^+}\approx n_{e^-}\approx 10^{31}$ cm$^{-3}$. The
annihilation's cross-section $\sigma_1$ of order $10^{-24}$ cm$^2$,
particles' velocity of light velocity's order; consequently, time
of equilibrium establishment is of order
$$
\tau=\frac{1}{\sigma_1nc}=10^{-17}\mbox{ñåê}.
$$
Thus, $\tau$ is insignificant in comparison with $t=1$ sec. Full
equilibrium $e^+ +e^-\rightleftarrows 2\gamma$ is provided. ...''

Rather later, in 1980, possibly, under the influence of
calibration field theory's results this viewpoint has been stated
more carefully \cite{ZeldDolg}:

 ``The problem of thermodynamical equilibrium in the initial plasma is extremely important.
If in ordinary cases equilibrium is reached after the lapse of
sufficiently great time, here the situation is exactly opposite.
At great $t$ expansion velocity of Universe proves to be greater
than velocity of reactions between particles and equilibrium has
no time to establish. The older the world becomes, the more
nonequilibrium it proves to be. On the contrary, at small times
$t$ reactions between particles become very fast in consequence of
density's increase and elementary particles' gas, as a matter of
fact proves to be equilibrium. Let us illustrate it in detail.
Expansion velocity of Universe is $\dot{a}/a\sim 1/t$. From the
other hand, the velocity of thermodynamical equilibrium's
establishment is $\dot{n}/n \sim nv\sigma$, where $n$ -
concentration of particles, $v$ - their velocity, and $\sigma$ -
interaction cross-section. Equilibrium, which is violated by world's
expansion, has time to recover, if:
$$
nv\sigma t\geq1. \eqno{(2.3)}
$$
At $T\gtrsim m$ density of particles $n$ by value order is equal
to $n(t)\approx (tt_{Pl})^{-3/2}$, where $t_{Pl}=G^{1/2}\approx
10^{-43}$c - Planck time; the inverse value
$T_{Pl}=t^{-1}_{Pl}\approx 10^{19}$GeV is called Planck
temperature (or mass). If interaction of particles is described by
unified calibrating theory, then $\sigma \sim \alpha^2T^{-2}$; in
case when temperature is greater than intermediate bosons' masses,
$m_X\approx 10^{15}$ GeV. Connection constant $\alpha$ at that
constitutes by order of value $10^{-2}$. Since temperature depends
on time by means of law $T\approx (tt_{Pl})^{-1/2}$, condition
of equilibrium (2.3) is fair at $t>\alpha^{-4}t_{Pl}$. However at
farther increase of $t$ (and decrease of temperature) situation
becomes equilibrium again.... Thus, at very small times $t<t_{Pl}$
Universe, possibly, is equilibrium; then at $t_{Pl}<t\lesssim
\alpha^{-4}t_{Pl}$ there is a nonequilibrium period and at
$\alpha^{-4}t_{Pl}>t>t_1$\footnote{Here inequalities should be
replaced by inverse ones, but this is a quotation. (Yu.I.)}
equilibrium recovers again... ''

\section{The Instructive Example}
Let us consider in detail given in book \cite{Zeld} example with
annihilation reaction. But for the estimation of LTE establishment
condition we will use not numerical values of parameters, given in
this book, but theirs analytic values. The total cross-section of
electron-positron pair's annihilation reaction is equal to (see
for example, \cite{Ahi})\footnote{Here and farther we choose the
universal system of units: $G=\hbar=c=k=1$, $k$ - Boltzmann
constant.}:
\begin{equation}\label{ann_s}
\sigma=\pi r^2_0\frac{\alpha^2}{4v_0\varepsilon^2_0}
\left\{\frac{3-v^4_0}{v_0}\ln\frac{1+v_0}{1-v_0}+2(v^2_0-2)\right\},
\end{equation}
where $\alpha=e^2/4\pi$ - constant of fine structure,
$\varepsilon_0$ - energy of colliding particles in c.m. system,
$v_0$ - their velocity in the same system. In particular, at
ul\-t\-ra\-re\-la\-ti\-vis\-tic energies of particles
$\varepsilon\gg m$; $v_0\to 1$ formula (\ref{ann_s}) gives:
\begin{equation}\label{ann_es}
\sigma=\pi
\frac{\alpha^2}{\varepsilon^2_0}\left(\ln\frac{2\varepsilon_0}{m}-1\right).
\end{equation}
Since for ul\-t\-ra\-re\-la\-ti\-vis\-tic particles $\varepsilon
\sim t^{-1/2}$, then substituting this relation into
(\ref{ann_es}), and after that into (6.2.8), we will obtain
instead of (6.2.9) the inverse inequality:
\begin{equation}\label{yu_ann}
t>t_1,
\end{equation}
- i.e. for the annihilation reaction LTE is absent in early times,
and recovers in later times. Is it possible that numerical
estimates, given in \cite{Zeld} are incorrect? Undoubtedly, these
estimates are correct, but at their finding there was implicitly
supposed, that in point of time $t=1$c:
\begin{enumerate}
 \item the number of electron-positron pairs in plasma did not differ
from the equilibrium value, de\-te\-r\-mi\-ned by temperature
$T(t)$ in given point of time;
\item plasma's temperature $T(t)$ in this point of time was determined
by formulas for locally equilibrium Universe, energy density in
which is proportional to $T^4(t)$.
\end{enumerate}

Thus, in book \cite{Zeld}, like, nevertheless in many others, has
been given an interesting result, which can be expressed by
following logical formula:

 {\it If in early Universe LTE existed, then LTE existed}!

The logical fault of such a conclusion is obvious. But in spite of
obvious falseness, this conclusion: ``{\it on early stages of
Universe there existed an LTE, which has been broken on later
ones}'' - was the foundation for construction of cosmological
evolution ideology, which with one or another variations in time
is called ``{\it the standard cosmological scenario}''.

\section{The Ideology Of Standard Cosmological Scenario And Its Consequences}
Let us consider basic features of standard cosmological scenario
(SCS), not wounding for the time being pro\-b\-lems of Universe
stability theory and connected with it pro\-b\-lems of
large-\-scale structure's formation.
\subsection{Space Homogeneity And Isotropy}
The first important statement of standard cos\-mo\-lo\-gi\-cal
scenario, as well as of overwhelming majority of cosmological
models, is the supposition about ho\-mo\-ge\-nei\-ty and isotropy
of three-dimensional space, that leads to Friedman metric:
\begin{equation}\label{Freed}
ds^2=a^2(\eta)(d\eta^2-dl^2)=dt^2-a^2(t)dl^2,
\end{equation}
where:
\begin{equation}\label{Freed_a}
dl^2=d\chi^2 +\rho^2(\chi)(d\theta^2+\sin^2\theta
d\varphi^2),\end{equation}
$$\rho(\chi)=\left\{\begin{array}{ll}
\sinh(\chi),& k=-1;\\
\chi, & k=0;\\
\sin(\chi), & k=+1\\
\end{array}\right.,$$
$k$ - curvature index of three-dimensional space: $k=0$ for zero
three-\-dimensional curvature, $k=1$ - for constant positive
three-\-dimensional curvature, and $k=-1$ - for constant negative
three-\-dimensional curvature. As is well known, Friedman metric
allows rotation group $O(3)$, generated by three Killing vectors
$\xi^i$:
\begin{equation}\label{Vec_Kill}
O(3):\; \left\{\begin{array}{lll}
\stackunder{1}{\xi}^i&=&\delta^i_\varphi;\\
\stackunder{2}{\xi}^i&=&
\delta^i_\theta\sin\varphi+\delta^i_\varphi\cos\varphi\cot\theta;\\
\stackunder{3}{\xi}^i&=&
\delta^i_\theta\cos\varphi-\delta^i_\varphi\sin\varphi\cot\theta,\\
\end{array}\right.
\end{equation}
only two of which are linearly independent, such that:
\begin{equation}\stackunder{\xi}{L}g_{ij}=\xi_{(i,j)}=0.\end{equation}
Besides, metric (\ref{Freed_a}) allows spacelike Killing tensor
field $\xi_{ij}$:
\begin{equation}\label{Ten_Kill}
\xi_{ij}=a^2(\eta)(g_{44}\delta^4_i\delta^4_j-g_{ij})
\end{equation}
such that:
\begin{equation}\xi_{(ij,k)}=0,\end{equation}
and timelike vector of conformal motion:
\begin{equation}\label{Kill_conf}
\stackunder{4}{\xi}=\delta^i_4,
\end{equation}
such that:
\begin{equation}\label{xi_conf}
\stackunder{\xi}{L}g_{ij}=\xi_{(i,j)}=2\frac{a'}{a}g_{ij}.
\end{equation}
As is well known, tensor of energy-momentum succeeds symmetry of
space-time in consequence of chain of relations:
$$\stackunder{\xi}{L}g_{ij}=0\Rightarrow
\stackunder{\xi}{L}R_{ijkl}=0\Rightarrow$$ and Einstein equations:
$$\stackunder{\xi}{L}R_{ij}=0\Rightarrow\stackunder{\xi}{L}T_{ij}=0.$$
Therefore tensor of energy-momentum of Friedman Universe takes
algebraic structure of Friedman metric, i.e., the structure of EMT
of ideal isotropic liquid:
\begin{equation}\label{MET}
T^{ij}=(\varepsilon +p)u^iu^j-pg^{ij},
\end{equation}
where
\begin{equation}\label{u^i}
u^i=1/\sqrt{g_{44}}\delta^i_4
\end{equation}
 - velocity vector of matter,
$\varepsilon(\eta)$, $p(\eta)$ - its energy density and pressure.

Einstein equations at that are reduced to two in\-de\-pen\-dent
equations (see for example, \cite{Land0}):
\begin{equation}\label{Einst1}
\frac{1}{a^2}(\dot{a}^2+k)=\frac{8\pi}{3}\varepsilon;
\end{equation}
\begin{equation}\label{Einst2}
\dot{\varepsilon}+3\frac{\dot{a}}{a}(\varepsilon+p)=0,
\end{equation}
where differentiation by time t is denoted by point. If we know
the equation of state, i.e., the relation of form:
\begin{equation}\label{p(e)}
p=p(\varepsilon),
\end{equation}
then equation (\ref{Einst2}) is integrated in quadratures:
\begin{equation}\label{a(e)}
a=a(\varepsilon).
\end{equation}
Substituting the solution (\ref{a(e)}) in equation (\ref{Einst1}),
we will obtain closed differential equation of first order
relatively to $\varepsilon(\eta)$. In case of {\it barotropic}
state equation:
\begin{equation}\label{bar}
p=\varrho\varepsilon
\end{equation}
Einstein equations are easily integrated for early Universe
($t\rightarrow 0$), as is well-known the behavior of solutions in
this case does not depend on the curvature index $k$ (see for
example, \cite{Land0}) and does not differ from the behavior of
solutions for the space flat Universe ($k=0$):
\begin{equation}\label{k+1not0}
a=a_1 t^{2/3(\varrho+1)};\; \varepsilon=\frac{1}{6\pi
(\varrho+1)^2t^2}, \quad \varrho+1\not=0
\end{equation}
and at $\varrho=-1$ we obtain the inflationary solution:
\begin{equation}\label{inflation}
a=a_1e^{\Lambda t};\quad
\varepsilon=\frac{3\Lambda^2}{8\pi}=\mbox{const}.
\end{equation}

Efforts of great number of theorists are directed to the formation
of such field models, which ensure required state equation
management: inflation, secondary acceleration, dark matter etc.
The dynamics of Friedman Universe geometry is exhausted by this,
but the dynamics of matter in this Universe is not.

\subsection{LTE And Algebra Of Interactions}
The second important statement of SCS is the hypothesis about
initial thermodynamical equilibrium of Universe, what became the
governing factor in theory of hot Universe formation. Starting
from the modern state of Universe and turning back its history
with the account of Friedman solution, describing the homogenous
cos\-mo\-lo\-gi\-cal expansion, as well as taking into account the
conservation law of number of particles and energy, we come to the
stage of hydrogen's recombination, before which photons laid in
LTE state with electrons and ions\footnote{From the point of view
of which we have spoken above.}. Thus, in early stages of
cos\-mo\-lo\-gi\-cal expansion act laws of equilibrium
thermodynamics, which are completely determined by
local-\-equi\-lib\-ri\-um dis\-t\-ri\-bu\-ti\-on functions of
particles.

So, let reactions of following type run in plasma:
\begin{equation}\label{reaction}
\sum\limits_A \nu_A a_A \rightleftarrows
\sum\limits_{B}\nu'_{B}a'_{B}
\end{equation}
where $a_A, a'_{B}$ - sort of particles (name), $\nu_A, \nu'_{B}$
- theirs numbers in this reaction. Then local-\-equi\-lib\-ri\-um
dis\-t\-ri\-bu\-ti\-on functions have form, (see for example,
\cite{Yukin1}):
\begin{equation}\label{f_LTE}
\stackrel{0}{f}_a(x,p)=\left[ \exp
\left(\frac{-\mu_a+(u,p)}{T}\right)\pm 1\right]^{-1}
\end{equation}
where  $T(x)$ - temperature and $u_i(x)$ - unit timelike vector of
macroscopic velocity $(u,u)=1$, the same for all sorts of
particles $a$; $\mu_a(x)$ - chemical potentials, which are
described by series of equations of chemical equilibrium:
\begin{equation}\label{chem_eq}
\sum\limits_A \mu_A \nu_A=\sum\limits_{B}\mu'_{B}\nu'_{B},
\end{equation}
representing the system of linear homogenous algebraic equations
relatively to $\mu_a$.
If in $k$ reaction of type (\ref{reaction}) certain vector
currents, created by corresponding charges $q_A$ è $q'_{B}$, are
conserved, then for such reactions the conservation law of charge
is fulfilled:
\begin{equation}\label{conserv_q}
\sum\limits_A q_A\nu^K_A-\sum\limits_{B}q'_{B}\nu^{'K}_{B}=0.
\end{equation}
Algebra of interactions of elementary particles, i.e., in fact,
schemes of reactions (\ref{reaction}), allowed in one or another
field-theoretical model of particles' interactions, leads to
conservation laws of certain generalized currents. Actually,
algebra of interactions of elementary particles is determined by
integers $\nu^K_n$, which are equal to number of particles of sort
$n$, participating in reaction denoted by index $K$, i.e., by
matrix $||\nu^K_n||$. Let $N$- number of fundamental particles',
including also antiparticles, in concrete field-theoretical model.
Let us rewrite reactions (\ref{reaction}) in unified form:
\begin{equation}\label{uni_reaction}
\sum\limits_{A=1}^N \nu^K_A a_A=0; \quad (K=1,2,\ldots),
\end{equation}
where $\nu^K_A$ can take already any integer values: positive,
negative and zero. In any closed field theory should be:
\begin{equation}\label{Rank}
\mbox{rank} ||\nu^K_A||<N,
\end{equation}
in the opposite case there will be such particle, which can be
obtained from others in not a single reaction
(\ref{uni_reaction}), i.e., will not interact with others, that
right now withdraws it outside the limits of given field theory,
making the mentioned one non-closed. In consequence of
(\ref{Rank}) we always can choose $N$ numbers $G_A$,
simultaneously different from zero, such that:
\begin{equation}\label{G_A}
\sum\limits_{A=1}^N \nu^K_AG_A=0;\quad (K=1,2,\ldots).
\end{equation}
Let for definiteness
$$\mbox{rank} ||\nu^K_A||=r<N.$$
Then there exist $N-r$ linear-independent solutions (\ref{G_A}),
which we will denote by means of symbols $G^s_A$
($s=\overline{1,N}$) and call generalized charges. Since $\nu^K_A$
- integers, solution of equations (\ref{G_A}) can always be
represented in rational numbers. Therefore, multiplying equations
(\ref{G_A}) by appropriating multipliers, we always can express
theirs solutions in integers, i.e., integer values can be attached
to generalized charges. Thus, in any closed field theory we will
have corresponding laws of generalized macroscopic currents
\cite{Yukin1}:
\begin{equation}\label{J^i}
J^i_s=\sum\limits_{A=1}^N G^s_A\int\limits_{P(x)}p^if_AdP_a.
\end{equation}
Since conditions of chemical equilibrium (\ref{chem_eq}) further
take form, formally identical to equations (\ref{G_A}):
\begin{equation}\label{mu_A}
\sum\limits_{A=1}^N \nu^K_A \mu_A=0;\quad (K=1,2,\ldots),
\end{equation}
then solutions of these equations with accuracy to within
multiplier do not differ from the solution of equations
(\ref{G_A}):
\begin{equation}\label{mu-g}
\mu^s_A=\sigma G^s_A,
\end{equation}
where $\sigma$ - common multiplier for all particles. From this
it, for example, right now follows, that if certain generalized
currents are conserved (for example, the electric current), then
massless quantums' chemical potentials of such field are equal to
zero, and chemical potentials of corresponding charged particles
and an\-ti\-par\-tic\-les are equal by absolute value and have
opposite signs.

Further, in homogenous and isotropic Universe all thermodynamical
functions should depend only on time, and vector of macroscopic
velocity should be equal to (\ref{u^i}). Then:
\begin{equation}\label{f(t)}
\stackrel{0}{f}_a(t,p)=\left[\exp\left(-\lambda_a(t)+\frac{E_a(p)}{T(t)}\right)\pm
1\right]^{-1},
\end{equation}
where:
\begin{equation}\label{E_a}
E_a(p)=\sqrt{m^2_a+p^2},
\end{equation}
- energy of particles ($p^2=-g_{\alpha\beta}p^\alpha p^\beta$ -
three-dimensional momentum's square),
\begin{equation}\label{lambda}
\lambda_a(t)=\frac{\mu_a(t)}{T(t)},
\end{equation}
- reduced chemical potentials, which also satisfy system of
equations of chemical equilibrium
\begin{equation}\label{lambda_A}
\sum\limits_{A=1}^N \nu^K_A \lambda_A=0;\quad (K=1,2,\ldots).
\end{equation}
In consequence of homogeneity of Universe and its isotropy
conservation laws of generalized currents (\ref{J^i}) in metric
(\ref{Freed}) on account of (\ref{u^i}) take form:
\begin{equation}\label{J^4}
a^3(t)\sum\limits_A G_A \Delta n_A(t)=\mbox{const},
\end{equation}
where $\Delta n_A$ - difference of densities of particles and
antiparticles of sort ``$A$'' with generalized charge $G_A$.

\subsection{High Entropy}
The third important statement of SCS is the statement of high
value of specific entropy, falling at one baryon in modern
Universe. More precisely, speech is about relation of photons'
quantity to baryons. It is convenient to incorporate the inverse
value:
\begin{equation}\label{delta}
\delta_B=\frac{n_B}{n_\gamma}\approx 10^{-10}\div 10^{-9}\ll 1,
\end{equation}
where $n_B$, $n_\gamma$ - densities of number of baryons and
photons in modern Universe, correspondingly.

Equilibrium densities of particles' number, $\stackrel{0}{n}$,
entropy, $\stackrel{0}{s}$, and energy,
$\stackrel{0}{\varepsilon}$, for gas of massless particles are
equal (see for example, \cite{Land}):
\begin{equation}\label{n_0}
\stackrel{0}{n}=\frac{\rho}{2\pi^2}\int\limits_0^\infty
\frac{p^2dp}{e^{p/T}\pm 1}=\frac{\rho T^3}{\pi^2}g_n\zeta(3);
\end{equation}
\begin{equation}\label{s_0}
\stackrel{0}{s}=\frac{d}{dT}\frac{\rho}{3\pi^2}\int\limits_0^\infty
\frac{p^3dp}{e^{p/T}\pm 1}=\frac{2\pi^2\rho T^3}{45}g_e;
\end{equation}
\begin{equation}\label{E_0}
\stackrel{0}{\varepsilon}=\frac{\rho}{2\pi^2}\int\limits_0^\infty
\frac{p^3dp}{e^{p/T}\pm 1}=\frac{\rho\pi^2 T^4}{30}g_e,
\end{equation}
where $\rho$ - number of independent polarizations of (spin)
particle ($\rho=2$ - for photons and massless neutrino), $g_a$ -
statistical factor ($g_a=1$ - for bosons, for fermions: $g_n=3/4$,
$g_e=7/8$), sign ``+'' corresponds fermions, ``-'' - bosons,
$\zeta(x)$ - $\zeta$ - Riemann function.

The summary energy density of massless particles is equal to:
\begin{equation}\label{E}
\varepsilon=\sum\limits_a \stackrel{0}{\varepsilon}_a={\cal N}
\frac{\pi^2 T^4}{15},
\end{equation}
where
\begin{equation}\label{g_E}
{\cal N}=\frac{1}{2}\left[\sum\limits_B (2S+1) +
\frac{7}{8}\sum\limits_F (2S+1)\right]
\end{equation}
- effective number of particles' types ($S$ - particle's
spin)\footnote{In field models of interactions of type SU(5)
${\cal N}\sim 100\div 200$.}; summation is carried out by bosons
(B) and fermions (F), correspondingly. Then summary entropy
density is equal to:
\begin{equation}\label{S}
s=\sum\limits_a \stackrel{0}{s}_a={\cal N} \frac{4\pi^2 T^3}{45}
\end{equation}
Let us consider now ultrarelativistic particles (baryons,
leptons), laying in thermal equilibrium, rest mass of which is
different from zero. Since chemical potentials of particles and
antiparticles are equal by value and have opposite signs, we will
obtain an expression for difference of massive baryons (leptons)
of certain type:
$$\Delta \stackrel{0}{n}=$$
\begin{equation}\label{Dn}
=\frac{\rho}{2\pi^2}\int\limits_0^\infty
\left[\frac{1}{e^{-\lambda
+E(p)/T}+1}-\frac{1}{e^{\lambda+E(p)/T}+1}\right]p^2dp\,.
\end{equation}
Supposing:
\begin{equation}\label{lambda}
\lambda_A=\frac{\mu_A}{T}\ll 1
\end{equation}
and proceeding to limit $m\to 0$ in integrals of type (\ref{Dn}),
we will obtain, expanding these integrals in series by smallness
of $\lambda$:
\begin{equation}\label{Dn_Lambda}
\Delta \stackrel{0}{n}\approx \lambda
\frac{\rho}{\pi^2}\int\limits_0^\infty
\frac{e^{p/T}}{(e^{p/T}+1)^2}dp=\lambda \frac{T^3}{3},
\end{equation}
where for definiteness we have put $\rho=2$ (S=1/2).

Thus, using formulas (\ref{n_0}) è (\ref{Dn_Lambda}), we will
obtain the relation for equilibrium ratio of baryons' excess to
number of photons:
\begin{equation}\label{Dn_Lambda/n}
\delta_B=\frac{\Delta
\stackrel{0}{n}}{\stackrel{0}{n}_\gamma}=\lambda\frac{\pi^2}{6\zeta(3)}\quad
( \approx 1,369\lambda),
\end{equation}
-  an equilibrium relative excess of baryons, $\delta_B$,
practically coincides with their reduced chemical potential:
\begin{equation}\label{delta=lambda}
\delta_B \sim \lambda.
\end{equation}
Since according to (\ref{n_0}) and (\ref{s_0}) equilibrium density
of ultrarelativistic particles' entropy is proportional to
equilibrium density of particles' number - $\stackrel{0}{s}\sim
\stackrel{0}{n}$, in standard cosmological scenario there draws a
conclusion about smallness of particles' chemical potentials on
ultrarelativistic stage of evolution of Universe, i.e., about
striking high degree of charge symmetry of Universe in the
beginning of evolution:
\begin{equation}\label{lambda_0}
\lambda\sim 10^{-10}\div 10^{-9}\ll 1.
\end{equation}
\subsection{Far-reaching Consequences}

From this right away arises the idea, whether Universe was from
the very beginning completely charge sym\-met\-ri\-cal or not; and
small excess of baryons ($\sim 10^{-10}$) originated in
consequence of some mechanisms of spon\-ta\-ne\-ous symmetry
breaking, which could take place at superhigh energies of
interacting particles, greatly exceeding principal experimental
possibilities of mankind. Exactly such idea has been stated by
Sakharov \cite{Sahar}(1967) and after that developed as a theory
of baryogenesis on the basis of SU(5)-model in papers
\cite{Kuzmin} - \cite{Fry3}. Baryogenesis theory applied
sufficiently strict conditions on minimal mass values of
extra-massive X-bosons (see for example \cite{Weinberg2}):
\begin{equation}\label{m_x}
m_X\geq 10^{16}\mbox{Gev}.\end{equation}%
Later in more exact author's calculations this limitation has been
reduced in 1.5 order (\cite{Ignat4} - \cite{Yu-Khal2}):
\begin{equation}\label{m_x_Yu}
m_X\geq 5\cdot 10^{14}\mbox{Gev},\end{equation}%
however, this does not change the matter of fact - the standard
cosmological scenario establishes limits on parameters of one or
another field theory of fundamental interactions. We can recall a
whole series of such ``cos\-mo\-lo\-gi\-cal'' limitations on
elementary particles' masses (neutrino, hadrons, gravitino etc.)
and other constants of fundamental interactions, obtained on the
basis of standard cosmological scenario (see for example,
fore-quoted book \cite{Zeld}, which presents the peculiar
encyclopedia of such limitations) and evoked earlier the
enthusiasm of hot model's followers. In turn, the combination of
conceptions of thermodynamical equilibrium and singular initial
condition of Universe with classical Hawking's results of
particles' creation by singularities has led to outwardly alluring
idea of vacuum origin of Universe, thus having made the initial
phase of Universe absolutely rigid and non-alternative.

Let us note, that such tendency - the obtaining of
``cosmological'' limitations on parameters of fundamental
interactions on the basis of SCS's consequences is utterly
dangerous for the development of the theory of fun\-da\-men\-tal
interactions at high energies - theories of fundamental
interactions become hostages of phe\-no\-me\-no\-lo\-gi\-cal
equilibrium model of Universe! This situation is illustrated on
fig. \ref{Ris1} in form of vicious circle. Exactly this vicious
circle has led the modern cosmology to the ideological crisis,
when the hope of Higgs bosons has gone and has appeared the series
of new experimental data, refereing to the structure of Universe
and unam\-bi\-gu\-ous\-ly interpreted within the limits of SCS as
the appeal to reconsideration of field theories' fundamental
principles. Isn't it easier to reconsider the validity of
principles of SCS itself?

\begin{center}
\parbox{8cm}{%
\vskip 1pt
\refstepcounter{figure}\epsfig{file=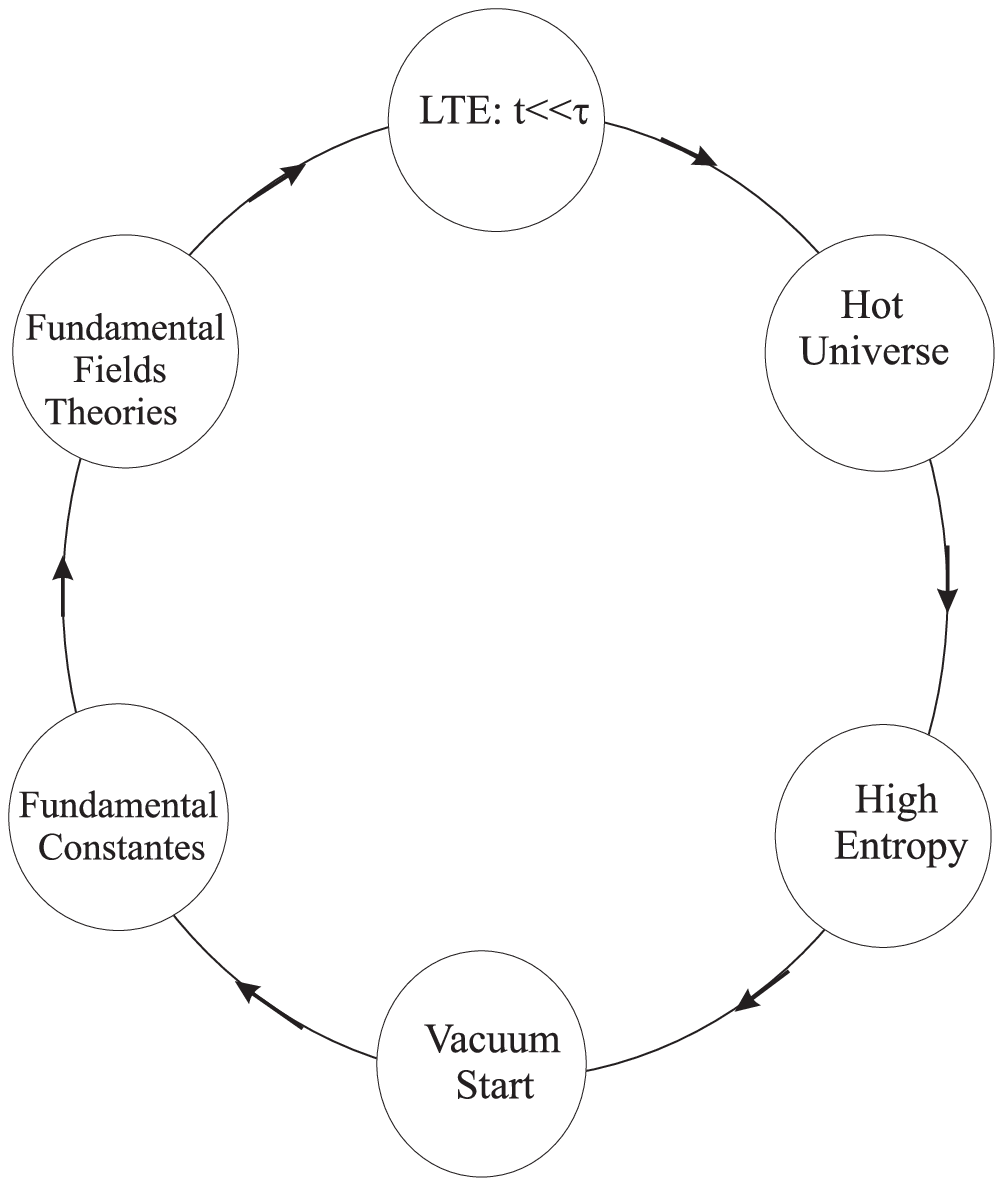,width=7cm,
height=8cm}\label{Ris1}
\vskip 1pt%
\hskip 1pt} \vskip 10pt\noindent Figure \thefigure:\quad The
Vicious
Circle Of Cosmology\end{center} %

\section{More Detailed Analysis Of LTE Conditions}
\subsection{The Influence Of Universe's singularity on LTE Establishment}

The first difference, which strikes our eyes at comparison of
cosmological process of LTE establishment with the ordinary
process - is the presence of the beginning of common history of
particles and their interactions in cosmology as against the
ordinary situation, which is induced by the presence of
cosmological singularity in point of time $t=0$. First, in
consequence of causality principle LTE can not be established in
times of Planck order or smaller. Actually, in sphere, formed by
light horizon of some one particle, is packed:
\begin{equation}\label{light_hor1}
N_t=\frac{4\pi}{3}t^3n(t)
\end{equation}
of other particles, where $n(t)$ - their number density. If
$N_t<1$, the interaction between particles can not take place, and
LTE will not establish. In hot model according to (\ref{k+1not0})
and (\ref{E}) ultrarelativistic plasma's temperature will vary by
law:
\begin{equation}\label{T_eq}
T_0(t)=\left(\frac{45}{32\pi^3{\cal
N}}\right)^{\frac{1}{4}}t^{-\frac{1}{2}},
\end{equation}
therefore at use of equilibrium concentrations (\ref{n_0}) the
relation (\ref{light_hor1}) in case of standard model SU(5) takes
form:
\begin{equation}\label{N_t}
N_t\sim 0,33 t^{\frac{3}{2}}.
\end{equation}
Thus, even at use of equilibrium concentrations of hot model LTE
can not be established at $t\lesssim t_{pl}$. But then initial
concentrations not in the least have to be equilibrium, - they can
turn to be greatly lower than last ones. But in this case the
establishment of LTE is put off for times later than Planck ones
\cite{Yudiffuz}.

Second, more detailed dynamical analysis of particles' correlation
functions also displays certain principal dif\-fe\-ren\-

ces of cosmological situation from the ordinary one. As the model
problem, which is solved exactly, we can consider the decay of
heavy electro-neutral rest massive particle into two
ultrarelativistic charged antiparticles (Fig. \ref{horizont}).

The exact solution of this problem is produced in A.V.\-Smirnov's
paper \cite{Smirnov} and is reduced to the substitution of kernel
$W_{ij}$ of relativistic integral of coulomb collisions of
Belyaev-Budker \cite{Belyaev} for the kernel $\bar{W}_{ij}$ by
rule:
$$\bar{W}_{ij}=W_{ij}\Theta(t),$$
where:
\begin{equation}\label{Smir}
\Theta(t)=\left\{%
\begin{array}{ll}%
0,&  0<t<\lambda_{pl}; \\
{\displaystyle \frac{1}{\Lambda}\ln \frac{t}{\lambda_{pl}}}, &
\lambda_{pl}< t
 < \lambda_D; \\
1, & t> \lambda_D,\\
\end{array}
\right.
\end{equation}
$\lambda_D$ - Debay-Hukkel radius, $\Lambda$ - coulomb logarithm.
This solution strictly shows, that particles' correlation before
Planck times is absent and only later on begin to increase
logarithmically slow up to classic relaxation times.
\vskip 12pt \noindent
\parbox{8cm}{%
\refstepcounter{figure}\begin{center}
\epsfig{file=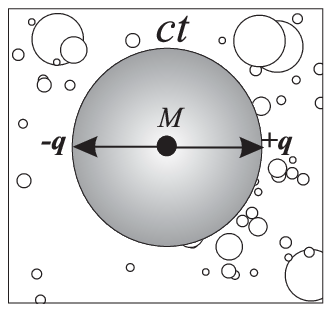,height=5cm,width=5cm}\end{center}\label{horizont}
} \vskip 10pt\noindent Figure \thefigure:\quad Light horizon of
two ultrarelativistic charges $q$ and $-q$, appeared at the decay
of electro-neutral massive particle $M$ in point of time $t=0$.
\vskip 12pt \hrule\vskip 12pt

\subsection{Conditions Of LTE}

Listed in section 3 principles of SCS are based on the condition
of LTE fulfilment in early Universe, - exactly this condition is
the main dogma of SCS. Therefore first of all we exactly should
check the fulfilment of this condition in early Universe, using
modern representations of elementary particles' interaction in
range of superhigh energies.

Since Universe's expansion temp is $\dot{a}/a$, more strict, than
(6.2.8) condition of LTE establishment has a form of:
\begin{equation}\label{LTE}
\tau_{eff}\frac{\dot{a}}{a}<1.
\end{equation}
If numbers of particles, participating in given reaction are
conserved:
\begin{equation} \label{yu1}
n(t)=\frac{n_1}{a^3(t)}, \end{equation}
where for definiteness we lay here and further:
\begin{equation}\label{yu2}
a(1)=1
\end{equation}
($t=1$ corresponds to planck point of time), $n_1=n(1)$-
particles' number density in this moment. According to
(\ref{Freed}) the choice of such normalization of scale factor
corresponds to the choice of planck length units in planck point
of time. At this normalization in case of barotropic equation of
state (\ref{bar}) we obtain from (\ref{k+1not0}):
\begin{equation}\label{yu3}
a=t^{2/3(\rho+1)};\quad \frac{\dot{a}}{a}=\frac{2(\rho+1)}{3t},
\quad(\rho\not= 1),
\end{equation}
and from (\ref{inflation}):
\begin{equation}\label{yu4}
a=e^{\Lambda(t-1)}; \quad \frac{\dot{a}}{a}=\Lambda,\quad
(\rho=-1).
\end{equation}

For more strict analysis of LTE condition it is necessary to
account the dependence of ultrarelativistic particles' interaction
cross-section $\sigma_{eff}$ from their kinetic energy $E_{cm}$ in
c.m.system. Since this energy is the function of cosmological
time, effective cross-section of scattering, as a matter of fact, is
also the function of time: $\sigma_{eff}=\sigma_{eff}(t)$.
Therefore condition of LTE (\ref{LTE}) takes form:
\begin{equation}\label{LTE1}
\frac{\dot{a}a^2}{n_1\sigma_{eff}(1)}\frac{\sigma_{eff}(1)}{\sigma_{eff}(t)}<1.
\end{equation}

For the clarification of function of effective interaction cross-section
from time it is necessary to consider in detail the kinematics of
four-particle reactions.

\subsection{The Kinematics Of Four-Particle Reactions And The Total Cross-Section Of Scattering}

Four-particle reactions of type:
\begin{equation}\label{Yu1}
a+b\rightarrow c+d
\end{equation}
are completely described by means of two kinematic invariants, $s$
and $t$, which posses following meaning: $\sqrt{s}$- energy of
colliding particle in center of mass (C.M.S):
\begin{equation}\label{Yu2}
s=(p_a+p_b)^2=(p_c+p_d)^2,
\end{equation}
and $t$-relativistic square of transmitted
momentum:\footnote{Author hopes that following notation's
coincidence will not confuse readers : t - time in Freedmans'
metric, s - its interval, simultaneously t, s - kinematic
invariants. This notation is standard and we didn't consider that
it is necessary to change it..}

\vskip 12pt \hrule \vskip 12pt \noindent
\parbox{8cm}{%
\refstepcounter{figure}\begin{center}
\epsfig{file=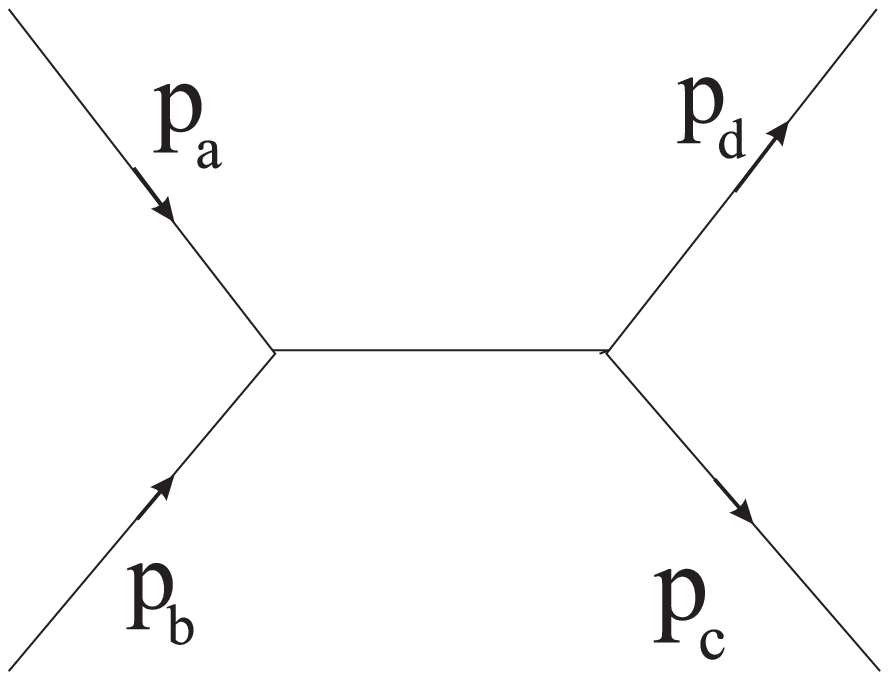,height=2.5cm,width=3cm}\end{center}\label{diagram}
} \vskip 10pt\noindent Figure \thefigure:\quad The diagram of
four-particle reaction \vskip 12pt \hrule

\begin{equation}\label{Yu3}
t=(p_c-p_a)^2=(p_b-p_d)^2,
\end{equation}
where momentum squares are understood as scalar four-piece
squares:
$$p_a^2=(p_a,p_a)=(p^4)^2-(p^1)^2-(p^2)^2 -(p^3)^2=m^2_a,$$
etc. For example:
$$(p_a+p_b)^2=p_a^2+2(p_ap_b)+p_b^2=m_a^2+2(p_a,p_b)+m_b^2.$$

Invariant scattering amplitudes $F(s,t)$, determined as a result
of averaging-out of invariant scattering am\-p\-li\-tu\-des by
particles state, $c$ and $d$, turn out to be depending only on
these two invariants (e.g., see \cite{Pilk}):
\begin{equation}\label{Yu4}
\sum |M_{FJ}|^2=\frac{|F(s,t)|^2}{(2S_c+1)(2S_d+1)},
\end{equation}
where $S_i$ - are spins. Using invariant amplitude $F(s,t)$ full
cross-section of reaction is determined (\ref{Yu1}) (see \cite{Pilk}):
\begin{equation}\label{Yu5}
\sigma_{tot}=
\frac{1}{16\pi\lambda^2(s,m^2_a,m^2_b)}\int\limits_{t_{min}}^0 dt
|F(s,t)|^2,
\end{equation}
where $\lambda$ - triangle function:
$$\lambda(a,b,c)=a^2+b^2+c^2-2ab-2ac-2bc;$$
$$t_{min}=-\frac{\lambda^2}{s}.$$
In ultrarelativistic limit:
\begin{equation}\label{Yu6}
\frac{p_i}{m_i} \to\infty
\end{equation}
we have:
\begin{equation}\label{Yu6_1}
s\rightarrow 2(p_a,p_b); \quad t\rightarrow -2(p_a,p_b),
\end{equation}
\begin{equation}\label{Yu6_2}
\frac{s}{m^2_i}\to\infty; \quad \lambda\to s^2,
\end{equation}
and formula (\ref{Yu5}) is considerably simplified by introduction
of dimensionless variable:
\begin{equation}\label{Yu7}
x=-\frac{t}{s}:
\end{equation}
\begin{equation}\label{Yu8}
\sigma_{tot}(s)=\frac{1}{16\pi s}\int\limits _{0}^1 dx|F(s,x)|^2.
\end{equation}
Thus, in ultrarelativistic limit total cross-section of scattering
depends only from the kinematic invariant $s$ - square of energy
of colliding particles in c.m.system:
$$\sigma_{tot}=\sigma_{tot}(s).$$
Exactly this dependence will manage the establishment of local
thermodynamical equilibrium in early Universe.

\subsection{The Influence Of Interaction Cross-section's Dependence
From The Kinematic Invariant $s$ Upon LTE Establishment}
Supposing henceforth the effective interaction cross-section is equal to
total one, we will investigate the dependence of LTE establishment from
the form of function $\sigma_{tot}(s)$. Let us suppose that for
ultrarelativistic particles there exists the power dependence of total
cross-section of scattering from the kinematic invariant (see \cite{Yudiffuz}):
\begin{equation}\label{Yus_1}
\sigma_{tot}(s)\sim s^\alpha, \qquad \alpha=\mbox{Const}.
\end{equation}
Since in isotropic expanding Universe the integral of motion is modulus
{\it of conformal momentum} of particle $\mathbb{P}$:
\begin{equation}\label{Yus_2}
a(t)p=\mathbb{P}=\mbox{Const},
\end{equation}
in ultrarelativistic limit (\ref{Yu6}) according to
(\ref{Yu6_1}) and (\ref{Yus_2}) the asymptotic behavior of kinematic
invariant is described by expression:
\begin{equation}\label{Yus_3}
\frac{p}{m}\rightarrow \infty \Rightarrow s\rightarrow
\frac{s_1}{a^2(t)},
\end{equation}
where $s_1=s(1)$. Thus, according to (\ref{Yus_1}) we will obtain:
\begin{equation}\label{Yus_4}
\sigma_{tot}(t)=\sigma_{tot}(1)a^{-2\alpha}(t).
\end{equation}

Substituting this dependence into LTE condition (\ref{LTE1}), we will obtain the explicit dependence of LTE condtion from the scale factor:
\begin{equation}\label{Yus_5}
\dot{a}a^{2(1-\alpha)}<n_1\sigma_{tot}(1).
\end{equation}
In that way, using solutions of Einstein equations for early Universe
in case of barotropic equation of state ($\rho\not=
-1$), (\ref{yu3}), we will obtain from (\ref{Yus_5}) the condition of LTE in early Universe:
\begin{equation}\label{Yus_6}
t^{ [4\alpha+3(1-\rho)]/(1+\rho)}<n_1\sigma_{tot}(1),
\end{equation}
from which follows, that at fulfilment of condition:
\begin{equation}\label{Yus_7}
4\alpha+3(1-\rho)>0,
\end{equation}
LTE is maintained in early stages of expansion, and is violated in late, i.e., at:
\begin{equation}\label{Yus_8}
\alpha > -\frac{3}{4}(1-\rho)  \Rightarrow \; LTE:\; t<t_0,
\end{equation}
and at fulfilment of inverse to (\ref{Yus_8}) condition LTE is violated in early
stages and is recovered in later stages. In case of ultrarelativistic equation of state $\rho=1/3$
we will obtain from (\ref{Yus_8}) the condition of existence of LTE in early stages \cite{Yudiffuz}:
\begin{equation}\label{Yus_8a}
\alpha > -\frac{1}{2} \Rightarrow \;LTE:\;
t<t_0,\quad(p=\frac{1}{3}\varepsilon).
\end{equation}
In case of extremely rigid equation of state $\rho=1$ the condition of LTE maintenance in
early stages and violation in later is equilvalent to condition:
\begin{equation}\label{Yus_8b}
\alpha > 0 \Rightarrow \;LTE:\; t<t_0,\quad(p=\varepsilon).
\end{equation}
In particular, at $\alpha=0$ (the interaction cross-section is constant) in case of
extremely rigid equation of state time at all comes out from the condition of LTE \cite{Yudiffuz},
 - in this stage of expansion either LTE in universe is always maintained or it is absent at all.
In case of inflation $\rho=-1$ LTE condition (\ref{Yus_6})
shoul be substituted for the following:
\begin{equation}\label{Yus_6a}
e^{\Lambda(3+2\alpha)(t-1)}<n_1\sigma_{tot}(1),
\end{equation}
therefore at:
\begin{equation}\label{Yus_9}
\alpha >-\frac{3}{2}
\end{equation}
LTE is maintained in early stages and is violated in later ones. {\it
Thus, the dependence of total cross-section of particles' interaction from
the kinematic invariant $s$ in range of superhigh values of energy plays the key role at clarification
of problem of LTE existence in early Universe}.

\section{Scaling Of Relativistic Particles' Interaction}
\subsection{Limitations On Scattering Cross-section's Asymptotic Behavior, Following From
The Axiomatic Theory Of S-Matrix}

There arises the question, what is the relation $\sigma_{tot}(s)$ in reality?
For analysis of kinetics of processes in early Universe it is
necessary to know the asymptotic behavior of invariant amplitudes
$F(s,t)$ in limit (\ref{Yu6}). Modern experimental opportunities
have coefficient restriction $\sqrt{s}$ at degree of hundreds GeV.
It would be risky to bear on that or other field model of
interaction for prediction of asymptotic behavior of scattering
crossection in the range of superhigh energies. It is more
rational in recent conditions to bear on axiomatic theory of
$S$-matrix conclusions get on basis of fundamental laws of
unitarity, causality, scale invariance etc. Unitarity of
$S$-matrix leads to well-known asymptotic relation (see, e.g.,
\cite{Okun})::
\begin{equation}\label{As1}
\left.\frac{d\sigma}{dt}\right|_{s\to\infty}\sim \frac{1}{s^2}
\end{equation}
for variables $s$ higher than unitary limit, i.e., under the
condition (\ref{Yu6}),if $m_i$ means the masses of all
in\-ter\-me\-di\-ate particles. But from (\ref{Yu8}) results:
\begin{equation}\label{As2}
F(s,1)|_{s\to\infty}\sim \mbox{Const}. \end{equation}
In the sixties of XX century on basis of axiomatic theory of
$S$-matrix were received stringent restrictions of asymptotic
behavior of total crossections and in\-va\-ri\-ant scattering
amplitudes:

\begin{equation}\label{As3}
\frac{C_1}{s^2\ln s}< \sigma_{tot}(s)<C_2\ln^2 s,
\end{equation}
where $C_1,C_2$ - unknown constants. Upper limit (\ref{As3}) was
determined in works \cite{asw1}-\cite{asw3}, lower limit - in
\cite{asw4}, \cite{asw5} (see also review in book \cite{asw6}). We
also notice restriction to invariant scattering amplitudes (see,
e.g., \cite{asw6}):
\begin{equation}\label{As4}
|F(s,t)|\leq |F(s,0)|;
\end{equation}
\begin{equation}\label{As5}
C'_1 < |F(s,0)|<C'_2 s\ln^2s.
\end{equation}

Therefore, invariant scattering amplitudes in limit (\ref{Yu6})
must be functions of variable $x=-t/s$, i.e.:

\begin{equation}\label{Yu15a}
|F(s,t)|=|F(x)|, \; (s \to\infty).
\end{equation}
But in consequence of (\ref{Yu8})
\begin{equation}\label{Yu15b}
\sigma_{tot}(s)=\frac{1}{16\pi s}\int\limits_0^1 dx
|F(x)|^2=\frac{\mbox{Const}}{s},-
\end{equation}
the total cross-section behaves itself such as the cross-section of
electromagnetic interaction, i.e. scaling is recovered at superhigh
energies.

Scaling asymptotics of cross-section (\ref{Yu15b}) lies strictly
between possible extreme asymptotics of complete scat\-te\-ring
cross-section (\ref{As3}). Moreover, at fulfilment of
(\ref{Yu15b}) relations, obtained on basis of axiomatic theory of
$S$-matrix (\ref{As1}) and (\ref{As2}) are automatically realized.

Further, as described above, scaling exists for pure
electromagnetic in\-te\-rac\-ti\-ons in consequence of their scale
invariance. As an example we will consider the annihilation cross-section
of ultrarelativistic electon-positron pair (\ref{ann_es}),
which can be rewritten with the help of kinematic invariant $s$ in obviously scaling form:
\begin{equation}\label{Yu15c}
\sigma_{ee\to
\gamma\gamma}=\pi\frac{\alpha^2}{s}(\ln\frac{2s}{m}-1).
\end{equation}

For lepton-hadron interaction assumption of scaling
existence has been offered in works \cite{scal1},\cite{scal2}.
In particular, for total cross-section of reaction
$$e+e^+\rightarrow \mbox{hadrons}$$
the following expression has been obtained:
$$\sigma_{tot}=\frac{4\pi\alpha^2}{3s}\sum e^2_i,$$
where $\alpha$ - fine structure constant, $e_i$ - charges of
fundamental fermion fields. Data, received on Stanford
accelerator, verify existence of scaling for this
in\-te\-rac\-ti\-ons. Apparently, for gravitational
in\-te\-rac\-ti\-ons scaling also must recover under superhigh
energies in con\-se\-qu\-en\-ce of scale invariance of
gravitational interactions in WKB-approximation \cite{scal4}. A
great number of analogous exam\-p\-les, presenting surely
established facts, can be given.

\subsection{The Asymptotic Conformal Invariance Of Relativistic Kinetic Theory}

The question is which consequences about ther\-mo\-dy\-na\-mi\-cal
equilibrium's establishment in early Universe gives us the strict
relativistic kinetic theory? Relativistic kinetic equations
relatively to macroscopic distribution function $f_a(x^i,p^k)$ of
particles of sort $a$ \cite{Yukin1} are,
\cite{Yukin4}-\cite{Yuconf}:
\begin{equation}\label{Yu00}
p^i\tilde{\nabla}_i f_a(x,p)=\sum\limits_{b,c,d}^{}
J_{ab\leftrightarrows cd}(x,p),
\end{equation}
where $\tilde{\nabla}$ - operator of covariant Cartan differentiation in phase space
$X\times P$:
\begin{equation}\label{Yu01}
\tilde{\nabla}=\nabla_i+\Gamma^j_{ik}p^k\frac{\partial }{\partial
p^j}.
\end{equation}

By means of distribution function $f_a(x,p)$ mac\-ro\-s\-co\-pic
moments are defined:
\begin{equation}\label{n_i}
n_a^i(x)=\int\limits_{P(x)}f_a(x,p)p^idP,
\end{equation}
- number of particles' of sort $a$ flux density vector and
\begin{equation}\label{T_ik}
T_a^{ik}(x)=\int\limits_{P(x)}p^ip^kf_a(x,p)dP,
\end{equation}
- particles' of sort $a$ tensor of energy-momentum, where
\begin{equation}\label{dP}
dP=\sqrt{-g} d^3p/p^4
\end{equation}
- invariant differential of momentum space's volume. Swinging the formula (\ref{T_ik})
by means of metric tensor $g_{ik}$, in consequence of normalization relation of 4 - momentum:
\begin{equation}\label{p^2}
(p,p)=m_a^2,
\end{equation}
we will obtain:
\begin{equation}\label{T_S}
T_S^a(x)=m^2_a\int\limits_{P(x)}f_a(x,p)dP,
\end{equation}
where $T_S^a(x)$ - spur of particles' of sort $a$ tensor of energy-momentum.

In case of homogenous isotropic distribution $f(\eta,p)$ in Friedman metric (\ref{Freed_a}
kinetic equations take form:
\begin{equation}\label{Yu03}
 \frac{\partial f_a}{\partial t}-\frac{\dot{a}}{a}
p\frac{\partial f_a}{\partial
p}=\frac{1}{\sqrt{m_a^2+p^2}}\sum\limits_{b,c,d}^{}
J_{ab\leftrightarrows cd}(t,p),\end{equation}
or in variables $\eta, p$:
\begin{equation}\label{Yu03a}
 \frac{\partial f_a}{\partial \eta}-\frac{a'}{a}
p\frac{\partial f_a}{\partial
p}=\frac{a(\eta)}{\sqrt{m_a^2+p^2}}\sum\limits_{b,c,d}^{}
J_{ab\leftrightarrows cd}(t,p),\end{equation}
where by means of a point is denoted the derivative by time $t$, and by means of a stroke -
the derivative by time variable $\eta$, at that:
$$a(\eta)d\eta=dt.$$
In paper \cite{Yuconf} it has been proved the following theorem:  {\it In ultrarelativistic
limit under the conditions of non - gravitational macroscopic field equations'
conformal invariance and scale invariance of matrix elements of interaction kinetic equations
are conformally invariant}.

Aforesaid means literally the following:
\begin{itemize}
\item Let us consider two conformally corresponding metrics
in common coordination:
\begin{equation}\label{Yu_Conf1}
d\bar{s}^2 =\sigma^2ds^2.\end{equation}
\item We will suppose, that at such transformation potentials
of scalar and vector fields are transformed by rule:
\begin{equation} \label{Yu_Conf2}
\bar{\Phi}=\Phi/\sigma; \qquad \bar{A}_i=A_i+\partial_i\varphi,
\end{equation}
where $\varphi(x)$ - scalar function, chosen in such way, that
calibration condition, imposed on vector po\-ten\-ti\-al stays
changeless at conformal transformation.
\item We will suppose, that at such transformation
canonical generalized momentums of particles, $P$, are transformed by law:
\begin{equation} \label{Yu_Conf3}
\bar{P}_i=P_i-\partial_i\varphi; \qquad
\bar{A}_i=A_i+\partial_i\varphi.
\end{equation}
\item Let us suppose, that at conformal transformation
in ultrarelativistic limit, when characteristic scales of the
system are smaller than compton scales of particles, all field
equations for non-gravitational interactions are asymptotically
con\-for\-mal\-ly in\-va\-ri\-ant\footnote{What corresponds to
WKB-approximation.} and matrix elements of interaction change only
in the issue of phase space's transformation at conformal
transformations:
\begin{equation}\label{Yu_Conf4}
\overline{|M(p,q|p',q')|^2}=\sigma^2(\eta)|M(p,q|p',q')|^2
\end{equation}
\item Then in ultrarelativistic limit the integral of pair collisions in
kinetic equations is transformed by law:
\begin{equation}\label{Yu_Conf5}
\bar{I}_{ab\leftrightarrow
cd}(\bar{P}_a)=\sigma^{-2}I_{ab\leftrightarrow cd}(P_a);
\end{equation}
\item The left side of kinetic equations (\ref{Yu03}) at that is transformed accurate within
members $O(p^2/m^2)$ by law:
\begin{equation}\label{Yu_Conf6}
\bar{\cal K}(x,\bar{P})= \sigma^{-2}{\cal K}(x,P),
\end{equation}
where ${\cal K}(x,P)$ - operator in the left side of kinetic equations.
\item Thus, if $f(x,P)$ - the solution of the kinetic equation in metric $g_{ij}$,
$f(x,\bar{P})$ in ultrarelativistic limit will be the solution
of kinetic equation in conformally corresponding metrics.
\end{itemize}

Using the conformal invariance of kinetic equations and the fact,
that, firstly, Friedman Universe (\ref{Freed}) at $k=0$ is
conformally -\-flat with the conformal multiplier
$\sigma=a(\eta)$, and, secondly, that in early stages of
cosmological expansion, when $\eta \rightarrow 0$, Friedman metric
tends asym\-p\-to\-ti\-cal\-ly to the space-flat, regardless of
the curvature index of three-dimensional space, $k$. Therefore
according to aforecited theorem the solution of kinetic equations
in metric (\ref{Freed}) will coincide with solutions of
corresponding kinetic equations in flat space,
$f_a(\eta,\bar{P}_a)$, where in corresponding kinetic equations it
is necessary to carry out a substitution of kinematic invariant
$s$ for $\bar{s}$ by rule:
$$\bar{s}=a^2(\eta)s, $$
i.e., $\bar{\sigma}_{tot}=\mbox{Const}$. But in that case $\bar{\tau}_{eff}=\mbox{Const}$,
and we come to the well-known result of standard kinetic theory: LTE is recovered in plasma at
$$\eta \geq \bar{\tau}_{eff}.$$
Thus, strict conclusions of relativistic kinetic
theory relatively to the recovery of LTE in ultrarelativistic plasma
are in full correspondence with the qualitative conclusion, given in the previous section.

\subsection{The Universal Asymptotic Cross-Section Of Scattering}

Further we will suppose the presence of scaling at energies above the unitary limit $s\to\infty$.
There arises the question about the meaning of constant in formula (\ref{Yu15b})
and also about the logarithmic correction of this constant.
This value can be estimated with the help of simple considerations.
If the idea of association of all interactions on Planck scales of energy $E_{pl}=m_{pl}=1$,
is correct, then at $s\sim 1$ all interactions should be described
by the united scattering cross-section, produced from three fundamental constants<
$G,\hbar,c$, i.e., in chosen system of units should be:
\begin{equation}\label{15c}
\sigma|_{s\sim 1}=\pi l^2_{pl}\quad (=\pi).
\end{equation}
However in order that on Planck scales of energy scattering
cross-section could fall up to such value, starting from the
values of order of $\sigma_T=8\pi\alpha^2/3m^2_e$ ($m_e$ -
electron's mass, $\sigma_T$ - Tompson scattering cross-section)
for elec\-t\-ro\-mag\-ne\-tic interactions, i.e., at $s \sim
m^2_e$, it should fall in inverse proportion to $s$, i.e., but
again by scaling law.\footnote{Let us note, that this fact is one
more independent argument in favour of scaling's existence in
range of high energies.} Improving this dependence
logariphmically, we will incorporate {\it The Universal Asymptotic
Cross-section Of Scattering} (ACS), introduced in papers
\cite{ACS}, \cite{Yuneq}:
\begin{equation}\label{Yu15d}
\sigma_0(s)=\frac{2\pi}{s\left(1+\ln^2\frac{s}{s_0}\right)}=\frac{2\pi}{s\Lambda(s)},
\end{equation}
where $s_0=4$ - square of total energy of two colliding Planck masses,
\begin{equation}\label{yu15e}
\Lambda(s)=1+\ln^2\frac{s}{s_0}\approx \mbox{Const},
\end{equation}
- logarithmic factor.

Incorporated by formula (\ref{Yu15d}), cross-section of
scat\-te\-ring $\sigma_0$, ACS, possesses the series of
outstanding features:
\begin{enumerate}
\item
ACS,  is produced only from fundamental constants $G, \hbar, c$;
\item
ACS,  behaves itself in such way, that its values lie strictly in
the middle of possible extreme limits of cross-section's
asymptotic behavior (\ref{As3}), es\-tab\-li\-shed with the help
of the asymptotic theory of $S$-matrix;
\item
ACS,  to a logarithmic accuracy is scaling cross-section of scattering;
\item
For reactions of photon's scattering on non-\-re\-la\-ti\-vis\-tic
electron ($s=m^2_e$) formula (\ref{Yu15d}) gives $\sigma_0=4/3
\sigma_T \sim \sigma_T$;
\item
For electro-weak interactions ($s=m^2_W$, where $m_w$ - mass of intermediate
$W$-boson) at  $\sin\theta_W=0,22$ (see for example \cite{Okun}) we will obtain from (\ref{Yu15d})
$\sigma_0=0,78\sigma_W$, where $\sigma_W=G^2_F m^2_W/\pi$ - cross-section of
$\nu e$ - scattering with the account of intermediate $W$-boson;
\item
At Planck values of energy $\sigma_0(m^2_{pl})\approx
\sigma_{pl}$.
\end{enumerate}
These outstanding features of ACS and its values' ama\-zing
coincidence with well-known processes' cross-\-sec\-ti\-ons in the
huge range of energy values (from $m_e$ to $10^{22}m_e$) hardly
can be casual, that allows us to apply ACS further in a capacity
of sure formula for the asymptotic value of scattring
cross-sections for all interactions.

Let us note, that coincidence of large-scale behavior of
cross-sections of elementary particles' interactions in range of superhigh
energies with ACS does not mean, that the same coincidence will conserve in
small scales of energies. Cross-sections' local deviations from ACS
will necessarily take place in form of so-called resonances (see Fig.
\ref{Res}).

\REFigure{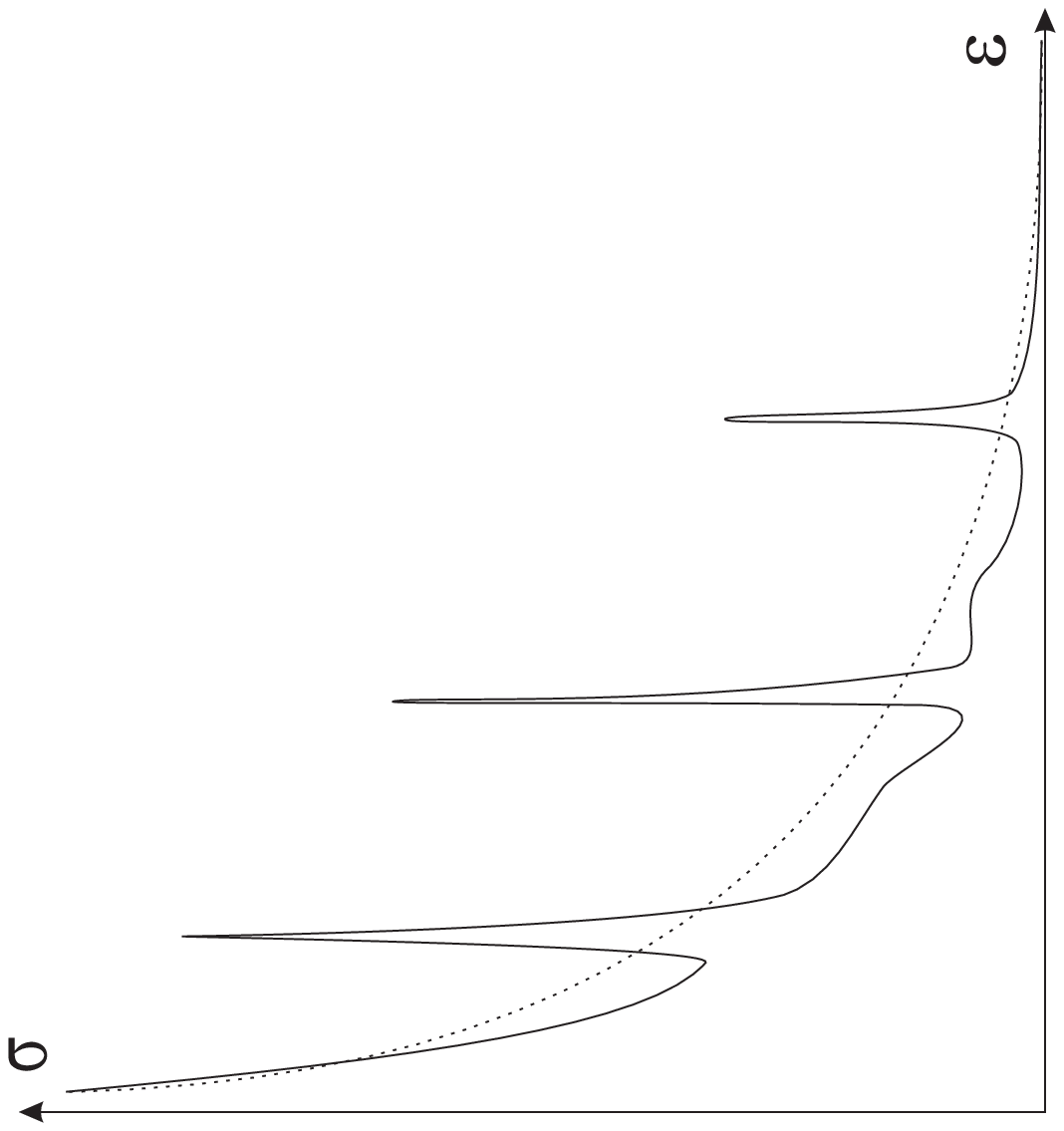}{The real dependence of total
cross-section of interaction on particles' energy. By means of dot
line is shown the behavior of ACS. Ranges of resonances in fact
can have much more complicated local structure. \label{Res}}

However, the influence of such resonances on general evolution
of cosmological plasma will disappear while energy of particles
\cite{Shul} will increase. Actually, masses of intermediate particles,
producing resonance, are of resonance's energy's order, which, in its turn,
is equal to the kinetic energy of interacting particles:
$$M\sim E \approx \sqrt{s}.$$
Therefore with the increase of the kinematic invariant $s$ increase also masses of
intermediate particles, determining resonances. But, as is well-known, there is the
following relation between resonance's half-width and mass of the intermediate
particle\footnote{This relation follows from the Geisenberg indeterminancy principle.}:
$$\Gamma\sim M^{-1}.$$
Therefore with the growth of energy resonances become
more and more narrow:
$$\Gamma \sim \frac{1}{\sqrt{s}},$$
such that their contribution in the kinetics of LTE establishment
becomes more and more weak.

Thus, summing up the results of the paper,
we can receive following consequences:
\begin{itemize}
\item The existence of LTE in early stages of Universe's evolution
is determined by the dependence of cross-sections of elementary particles' interactions
in the range of superhigh energies on kinematic invariant,
$s$, - energy of interacting particles in c.m.system; %
\item In case of ultrarelativistic equation of state of
cosmological plasma and power dependence of cross-section of interaction on
kinematic $\sigma\sim s^{\alpha}$ LTE is absent in early stages of Universe's expansion at
$\alpha<-1/2$;
\item
The quantum theory of field predicts the recovery of scaling of
interactions at superhigh energies of particles in consequence of
conformal invariance in ultrarelativistic limit of fundamental
field equ\-a\-ti\-ons. Elementary particles' cross-sections of
in\-te\-rac\-ti\-ons at that in the range of superhigh energies
are inverse in proportion to the kinematic invariant $s$;
\item
In conditions of scaling of interactions LTE should be violated in
early Universe and should be re\-co\-ve\-red in later stages;
\item
Since LTE is absent in early stages of Universe,
the initial distribution of particles can be random and can differ greatly from the
equilibrium one;
\item
Since all particles' interactions at superhigh ener\-gi\-es are
unified, interactions of all particles in this range can be
qualitative correctly  described by means of the universal
asymtotic cross-section of scattering, ACS, which possesses the
scaling character.

\end{itemize}

Since further we will investigate the kinetics of reactions only
in the range of superhigh energies, where all interactions are
described, as we suppose, by ACS, no difference can be made
between particles in integrals of interactions, accounting only
where it is essential, their spin and other characteristics. From
this point of view all interactions are unified at superhigh
energies, what greatly simplifies the investigation of such
processes. In the next article we will research the kinetics of
LTE establishment in the early Universe, using ACS for the
description of particles' interactions in the range of superhigh
energies, and also we will clarify the bo\-un\-da\-ri\-es of
arbitrariness of initial distribution of particles.
%
%


\end{document}